# Discovery of universal phonon thermal Hall effect in crystals

Xiaobo Jin[1*], Xu Zhang[1*✉], Wenbo Wan[1], Hanru Wang[1], Yihan Jiao[1] and Shiyan Li[1,2,3✉]

[1]*State Key Laboratory of Surface Physics, Department of Physics, Fudan University, Shanghai 200433, China*

[2]*Shanghai Research Center for Quantum Sciences, Shanghai 201315, China*

[3]*Collaborative Innovation Center of Advanced Microstructures, Nanjing 210093, China*

Corresponding author. Email: shiyan_li@fudan.edu.cn (S.Y.L.); xuzhang_fd@fudan.edu.cn (X.Z.)

## Abstract

**Thermal Hall effect (THE) in insulator is a remarkable phenomenon that arises from the motion of chargeless quasi-particles under a magnetic field[1-11]. While magnons[1-3] or exotic spin excitations[4,5,12,13] were considered as the origin of THE in some magnetic materials, there are more and more evidences suggesting that phonons play a significant role[8-11,14-23]. However, the mechanism behind phonon THE is still unknown. Here we report the observation of THE, including planar THE, in a broad range of non-magnetic insulators and semiconductors: $SrTiO_3$, $SiO_2$ (quartz), MgO, $MgAl_2O_4$, Si and Ge. While the presence of antiferrodistortive domains in $SrTiO_3$ and chiral phonons in $SiO_2$ may complicate the interpretation of THE, the striking observations of THE in trivial insulators MgO and $MgAl_2O_4$, as well as in high-purity intrinsic semiconductors Si and Ge, demonstrate that phonon THE is a universal property of crystals. Without other effects on phonons such as from magnons, this universal phonon THE is characterized by a scaling law of $|\kappa_{xy}| \sim \kappa_{xx}^2$. Our results experimentally discover a fundamental physics of phonons in magnetic field, which should come from the direct coupling between atom vibrations and the field. Starting from this universal phonon THE in crystals, all previous interpretations of THE in magnetic or non-magnetic materials need to be reconsidered.**

# Introduction

Hall effect as a fundamental phenomenon in condensed matter physics, plays a crucial role in investigating exotic quantum materials[24-27]. In conductors, it manifests as a transverse voltage difference caused by a longitudinal electrical current under a perpendicular magnetic field. However, in insulators, where the electrical Hall effect is ineffective due to the absence of charge carriers, the thermal Hall effect (THE) becomes important. THE arises from the motion of quasi-particles driven by a longitudinal temperature gradient in the presence of a magnetic field, leading to a transverse temperature difference. Over the past decade, non-zero thermal Hall conductivity $\kappa_{xy}$ has been detected in a few insulators[1-11], yet its nature remains an open question.

To date, three distinct interpretations of the THE in insulators have been proposed, including magnons, exotic magnetic excitations such as Majorana fermions, and phonons. First, in ferromagnetic insulators, the large THE is attributed to magnons with the nontrivial topology of magnon bands[1-3]. Second, beyond magnons, THE has also been considered to arise from exotic spin excitations in frustrated quantum magnets such as $Tb_2Ti_2O_7$ (ref. 4) and $\alpha$-RuCl$_3$ (refs. 5,12,13,28). Particularly in honeycomb-lattice $\alpha$-RuCl$_3$, the observation of half-integer quantized THE was argued to support the Majorana fermions origin[5,12,13]. Third, more and more evidences show that phonons play an important role in the observed THE across a wider range of materials, including the paramagnetic insulators[6,21], multiferroic materials[8], cuprates[7,14,15], antiferromagnetic (AF) insulators[9,16], Kagome materials[17,18], the metallic spin ice $Pr_2Ir_2O_7$ (ref. 19) and the van der Waals ferromagnet $CrI_3$ (ref. 20). Interestingly, planar THE (here, "planar" refers to the field lying in the $xy$ plane of the sample, as illustrated in Fig. 1a) has also been observed in several magnetic insulators with the origin related to phonons[22,23].

However, the mechanism for phonons to generate THE remains elusive. In the case of paramagnetic insulator $Tb_3Ga_5O_{12}$ (ref. 6), the phonon THE is linked to skew scattering with magnetic ions[29]. In magnetic insulators, the coupling of phonons to the magnetic environment is widely considered[8,9,14-16,30-34]. Interestingly, the phonon THE has also been reported in two non-magnetic insulators[10,11]. In SrTiO$_3$ (STO), THE is associated with antiferrodistortive (AFD) domains, which potentially lead to hybridization between acoustic and transverse optical phonons[10]. In black phosphorus (BP), the THE may stem from an intrinsic scenario where atomic vibrations can couple to the field due to the anisotropic charge distribution[11]. To further elucidate the essence of phonon THE in this context, it will be crucial to measure more non-magnetic insulators, to check how often it can be observed and whether planar THE also exists.

In this Article, we carry out systematic thermal Hall measurements on non-magnetic single crystals, including insulating SrTiO$_3$, SiO$_2$, MgO and MgAl$_2$O$_4$, as well as semiconducting Si and Ge. Firstly, we observe large conventional and planar $\kappa_{xy}$ in non-magnetic SrTiO$_3$ and SiO$_2$, which indicates that planar THE does not depend on magnetic environment and shares common nature with the conventional THE. Secondly, the large $\kappa_{xy}$ is also observed in trivial insulators such as MgO, MgAl$_2$O$_4$, and high-purity elemental semiconductors Si and Ge, which can be described by a scaling law of $|\kappa_{xy}| \sim \kappa_{xx}^2$. These findings demonstrate that phonon THE is a universal phenomenon in crystals (even with high purity), which reflects a fundamental physics of phonons (i.e., atom vibration) in magnetic field. Therefore, the interpretation of all previous THE results should start from an existing large intrinsic phonon THE, then consider other factors to affect this phonon THE, instead of seeking for some exotic origins such as Majorana fermions.

**Results**

We start with thermal Hall measurements on STO to investigate the potential existence of planar THE in non-magnetic insulator. The schematic diagram of experiment setup is shown in Fig. 1a. Its high credibility has been validated by measuring glass, Cu$_3$TeO$_6$, VI$_3$ and CrI$_3$, and comparing with those results in previous reports (more details are available in Supplementary Note 1 and 2). Figure 1b and 1c show the thermal transport results of STO. The behavior of both $\kappa_{xx}$ and conventional $\kappa_{xy}$ ($H \parallel z$) is consistent with the results in previous report[10]. Clearly, $\kappa_{xy}$ shows up in both planar configurations ($H \parallel x$ and $H \parallel y$). Although the magnitude of the planar $\kappa_{xy}$ is smaller than the conventional one, all three curves exhibit a peak around 20 K, following $\kappa_{xx}$ curve in Fig. 1b. In this non-magnetic material STO, phonon is the only well-defined quasiparticle, leaving no doubt that THE comes from phonons. Our results show that planar phonon THE also exists in non-magnetic material, without the need to interact with a magnetic environment.

We then direct our focus to the "chiral" crystal $SiO_2$, since the chirality of phonons is often believed necessary for phonon THE. $SiO_2$ is characterized by helical $SiO_4$ tetrahedra along [001] (see Supplementary Fig. S3). A recent study has detected chiral phonons carrying magnetic moment in $SiO_2$ using circularly polarized X-rays[35], which motivate us to check whether these chiral phonons will give a large phonon THE in $SiO_2$. As illustrated in Fig. 2b, $SiO_2$ show a significant $\kappa_{xy}$ in an out-of-plane field. The peak amplitude reaches approximately 1.8 W m$^{-1}$ K$^{-1}$, comparable to values reported in materials where the $\kappa_{xy}$ is considered "large", such as $Cu_3TeO_6$ (ref. 9) and BP[11]. Large planar $\kappa_{xy}$ with both $H \parallel x$ and $H \parallel y$ is also observed in $SiO_2$ with values exceeding half of the conventional $\kappa_{xy}$, further confirming our finding in STO.

Given the unique crystal structure for chiral phonons in $SiO_2$, one might not expect to observe the THE in a trivial insulator like MgO. Our motivation to measure MgO is to take it as a standard crystal without phonon THE. However, to our surprise, a large $\kappa_{xy}$ is also observed (Fig. 3b). It is evident that the shape of the $\kappa_{xy}(T)$ curve for MgO closely follows the phonon peak on its $\kappa_{xx}$ curve, resembling that of STO and $SiO_2$. It is important to note that MgO possesses a trivial crystal structure, which poses challenge to those interpretations related to domain in STO or chiral structure in $SiO_2$. To explain this phonon THE in MgO, we notice that Flebus and MacDonald recently proposed an intrinsic scenario emphasizing the influence of the Lorentz force on ionic vibrations without the coupling of acoustic phonons to extrinsic factors[36]. In this scenario, the in-phase motion of cations and anions couples with out-of-phase motion under magnetic fields, thereby generating finite thermal Hall signals.

If this ionic scenario of the phonon THE is true, one may not expect to observe THE in an elemental crystal. Therefore, we choose to measure the high-purity elemental intrinsic semiconductors Si. Strikingly again, a large $\kappa_{xy}$ is also observed in Si (Fig. 3d). Intrinsic Si single crystal has a 1.12 eV band gap. According to the Wiedemann-Franz law, the electron contribution to $\kappa_{xy}$ can be neglected at low temperature. Thus, the large $\kappa_{xy}$ must entirely come from phonons, and the effect of impurity is also negligible in this high-purity single crystal. Its peak magnitude is about 10 W m$^{-1}$ K$^{-1}$, which is the largest in all known crystals so far. The observations of THE in STO, SiO$_2$, MgO, and Si finally let us realize that the phonon THE is a universal phenomenon in any crystal, even with high purity like Si. In this context, special AFD domain like in STO or chiral structure like in SiO$_2$ may only slightly affect this phonon THE, and previous observation of large THE in elemental BP single crystal[11] is also not surprising now. More measurements on another insulator MgAl$_2$O$_4$ with spinel structure and intrinsic semiconductor Ge are carried out to reinforce this discovery. As expected, both single crystals display a large $\kappa_{xy}$, shown in Fig. 3. We note that previously two crystals were reported to have negligible or tiny $\kappa_{xy}$, namely Y$_2$Ti$_2$O$_7$ (ref. 4) and KTaO$_3$ (ref. 10). In fact, a non-zero $\kappa_{xy}$ in Y$_2$Ti$_2$O$_7$ has been observed, but it was attributed to temperature gradients on the Cu heat sink[4]. We conducted a comparative experiment on SiO$_2$ to investigate the effect of the Cu heat sink, and found no detectable contamination on thermal Hall signal from it (more details are available in Supplementary Note 4). Therefore, it can be inferred that the non-zero $\kappa_{xy}$ in Y$_2$Ti$_2$O$_7$ is intrinsic. The thermal Hall angle (tan$\theta$) of Y$_2$Ti$_2$O$_7$ is approximately $0.25 \times 10^{-3}$ at 15 K and 9 T, which is comparable to that of SrTiO$_3$. Similarly, previous study reported that KTaO$_3$ exhibits a tiny $\kappa_{xy}$ (ref. 10). We conducted a thermal Hall experiment on KTaO$_3$ and observed a peak value of $\kappa_{xy}$ approximately one-third of STO (Fig. S5). These findings further confirm the universal phonon THE in all crystals.

This universal phonon THE is characterized by a scaling law between $\kappa_{xy}$ and $\kappa_{xx}$, in case that no other extrinsic factors to affect phonons. The $\kappa_{xy}$ of SiO$_2$, MgO, Si and Ge is proportional to the square of $\kappa_{xx}$, as depicted in Figs. 4a-4d. The data of MgAl$_2$O$_4$ and the data of BP extracted from ref. 11 are plotted in Fig. S7, which also show a square dependence. Thus, a universal scaling law of $|\kappa_{xy}| \sim \kappa_{xx}^2$ for intrinsic phonon THE is presented in Fig. 4e. It can also be written as $\kappa_{xy}/\kappa_{xx} \sim \kappa_{xx}$, which means the Hall angle

tanθ = $\kappa_{xy}/\kappa_{xx}$ is determined by $\kappa_{xx}$. Note that for BP in ref. 11, $x$ and $z$ axes are defined in the plane and $y$ axis is defined perpendicular to the plane, therefore the universal scaling law is $|\kappa_{xz}| \sim \kappa_{xx}^2$. Scaling law is particularly relevant in the context of anomalous Hall effect studies, where distinct scaling laws originate from various mechanisms[24,37]. For instance, localized hopping is indicated by $|\sigma_{xy}| \sim \sigma_{xx}^{1.6}$, while skew scattering is represented by $|\sigma_{xy}| \sim \sigma_{xx}$. The scaling law in thermal transport reinforces our discovery regarding the universal phonon THE in crystals.

## Discussion

Revealing the origin of this universal THE in crystals requires a comprehensive understanding of the thermal Hall signal in elemental semiconductors Si and Ge. Even though the scenario of ionic vibrations[36] may explain the thermal Hall phenomena in $SiO_2$ and MgO, it falls short in explaining the case of elemental Si and Ge. For another elemental crystal BP, which also exhibits large $\kappa_{xy}$, it has been suggested that the unevenly distributed positive and negative charges in BP could pave the way for coupling between phonons and the magnetic field. However, the high symmetry of Si and Ge crystals poses challenge in forming such uneven charge distributions. Note again, the data of BP also exhibit the same scaling law indicating a common origin with our compounds.

Considering the high quality of our non-magnetic single crystals, especially the high-purity Si and Ge (with purity higher than 4N), extrinsic mechanisms for phonon THE such as coupling to the magnetic environment, impurity scattering, and AFD domains are unlikely. In this context, an intrinsic mechanism, i.e. the direct influence of Lorentz forces on the coupled phonon modes during phonon propagation, should exist for this universal phonon THE. Ceresoli *et al*. conducted calculations on the model of hydrogen molecules and revealed that the magnetic field effect survives for the relative motion of the two atoms, despite perfect screening for center-of-mass motion[38]. Previous studies have proposed Raman-type interactions between the magnetic field and phonons, leading to an intrinsic THE in non-magnetic band insulators[39-41]. However, in long-wavelength limit, the coupling of acoustic phonons to the magnetic field occurs through higher order gradient terms, resulting in an intrinsic effect much smaller than the measured values. In particular, the calculation of intrinsic phonon thermal Hall conductivity for Si gave a positive $\kappa_{xy} \sim 10^{-6}$ W m$^{-1}$ K$^{-1}$ at 300 K, which decreases with decreasing temperature[39], far away from our experimental result. Therefore, the microscopic mechanism for this universal phonon THE with a scaling law $|\kappa_{xy}| \sim \kappa_{xx}^2$ remains an open question. New theoretical approach and calculations are highly desired to clarify this fundamental physics of phonons in magnetic field.

Since the direct coupling between atom vibrations and the field gives this universal phonon THE, it explains why there exists planar phonon THE. In fact, previously the observation of planar THE in α-RuCl$_3$ (refs. 12,13,42), Na$_2$Co$_2$TeO$_6$ (ref. 23) and cuprates[22] was quite puzzling. Now this planar THE is also observed in our STO and SiO$_2$ samples. Furthermore, the planar $\kappa_{xy}$ curves of SiO$_2$ follow the same scaling law (see Supplemental Fig. S8), confirming that the planar THE originates from the same source as the conventional one. In three-dimensional crystals, phonon vibration contains two transverse waves and one longitudinal wave, allowing for the convenient coupling of a magnetic field in any direction to affect phonon vibrations. This ultimately manifests in the THE through coupling between phonon modes in any direction, including the planar configurations.

Starting from this universal phonon THE in crystals, all previous interpretations of THE in magnetic or non-magnetic materials need to be reconsidered. It is a conceptional change. Previously, people were trying hard to find extrinsic mechanisms for phonons to generate the THE, such as impurity scattering or interaction with the magnetic environment, since phonon itself is usually believed not affected by magnetic field. Now one must accept that phonon THE is a universal intrinsic property of any crystal. Bearing this in mind first, then one may consider various factors which will affect this phonon THE, such as magnetic environment, impurity, and doped carriers.

In ferromagnetic insulators such as Lu$_2$V$_2$O$_7$ (ref. 1), Fe$_2$Mo$_3$O$_8$ (ref. 8), VI$_3$ (ref. 3) and CrI$_3$ (ref. 20), the field-dependent $\kappa_{xy}$ tracks the magnetization curve. This means that the spin polarization affects the phonon $\kappa_{xy}$ significantly. We notice that $\kappa_{xy}$ of CrI$_3$ is negative and only changes slightly at $T_c$, roughly following the $\kappa_{xx}$ curve[20]. On the contrary, $\kappa_{xy}$ of $Lu_2V_2O_7$, $Fe_2Mo_3O_8$ and $VI_3$ is positive, and changes abruptly at $T_c$, not following the $\kappa_{xx}$ curve[1,3,8]. More works are needed to clarify the effect of spin polarization on the phonon THE.

In the quantum spin-liquid candidate $\alpha$-$RuCl_3$, initially half-integer quantized anomalous THE was reported in both conventional and planar configurations, which was argued as evidence for Majorana fermions[5,12]. However, this half-integer quantized anomalous THE is not robust, since it cannot be reproduced by some other groups[17,42]. Based on current work, we know that there must exist an intrinsic phonon THE in $\alpha$-$RuCl_3$, as reported in ref. 17 in a completely conventional configuration. Therefore, the observed field dependence of $\kappa_{xy}$ in $\alpha$-$RuCl_3$ around the critical in-plane field 7.5 T may only reflect the effect of magnetic fluctuations on the phonon THE, thus no need for introducing exotic Majorana fermions.

In cuprates, starting from this universal phonon THE, the observation of large $\kappa_{xy}$ in $La_2CuO_4$ (ref. 7), $Na_2CuO_4$ and $Sr_2CuO_2Cl_2$ (ref. 15) is not surprising now. While the energy gap of magnons in these AF insulators is large, short-range magnetic fluctuations may still affect this phonon THE. The main question in cuprates is why the phonon $\kappa_{xy}$ vanishes at high doping[14]. It is likely that the increasing carriers, as scatterers of phonon, gradually diminish the phonon THE. In another system $Sr_2Ir_{1-x}Rh_xO_4$, the intrinsic phonon $\kappa_{xy}$ is detected in the AF parent compound $Sr_2IrO_4$ (ref. 16). The huge change in $\kappa_{xy}$ emerging from element substitution on the spin-carrying site shows that the phonon THE is strongly affected by spin-phonon coupling and impurity scattering in this system[16]. All above mentioned factors will affect the phonon THE, therefore the scaling law $|\kappa_{xy}| \sim \kappa_{xx}^2$ may no longer hold. Indeed, as seen in Supplementary Fig. S9, most magnetic insulators do not obey this scaling law.

In summary, we have experimentally discovered a universal phonon THE in crystals, characterized by a scaling law of $|\kappa_{xy}| \sim \kappa_{xx}^2$, by measuring a series of non-magnetic insulators and intrinsic semiconductors, including STO, $SiO_2$, MgO, $MgAl_2O_4$, Si and Ge. It shows that phonons do not require extrinsic factors to generate THE. This is counter-intuitive since phonon itself is usually believed not affected by

field. This completely changes the starting point for interpreting the THE results in magnetic or non-magnetic materials. The nature behind this fundamental physics of phonons in magnetic field should relate to the direct coupling of atom vibrations to the magnetic field, which require further theoretical investigations.

# Methods

**Samples.** Single crystals of $SrTiO_3$, $SiO_2$, MgO, $MgAl_2O_4$, Ge and $CrI_3$ were purchased from Prmat (Shanghai) Technology Co., Ltd. The $KTaO_3$ single crystal was purchased from Semiconductor Wafer Inc. The high-purity intrinsic semiconductor Si single crystal was purchased from Bonda Technology Pte Ltd. The $Cu_3TeO_6$ single crystals were grown by Prof. Yuan Li's group[43]. The $VI_3$ single crystals were grown by Prof. Hechang Lei's group[44]. For thermal transport measurement, the samples were prepared in the shape of a rectangular platelet. The platelets of $SiO_2$ have the largest face perpendicular to the [110] axis and the longest edge along [001] axis. The largest faces of $SrTiO_3$, $KTaO_3$, MgO, $MgAl_2O_4$, Si and Ge are perpendicular to the [100] axis. The dimensions (length between contacts $L$ × width between contacts $w$ × thickness $t$) of all the measured samples in the main text are listed in Table 1.

| sample | $L$ (*mm*) | $w$ (*mm*) | $t$ (*mm*) |
| --- | --- | --- | --- |
| $SrTiO_3$ | 2.19 | 2.04 | 0.157 |
| $KTaO_3$ | 2.25 | 2.04 | 0.200 |
| $SiO_2$ | 2.64 | 1.32 | 0.086 |
| MgO | 2.43 | 2.07 | 0.100 |
| $MgAl_2O_4$ | 2.39 | 1.82 | 0.125 |
| Si | 2.50 | 1.25 | 0.086 |
| Ge | 1.93 | 1.68 | 0.129 |

**Table 1**: Dimensions (length between contacts × width between contacts × thickness) of samples used in this work.

**Measurements.** The thermal transport measurements were conducted in a Physical Property Measurement system (PPMS, Quantum Design) under high-vacuum environment. Three-thermometers (Cernox 1050) method was employed to simultaneously measure the longitudinal and transverse thermal gradients. The thermometers are connected to the sample by silver wires. The contacts were made of silver epoxy annealed at 380 K. A constant heat current $Q$ was applied along the $x$ axis lying in the basal plane of the single crystal, using a resistive heater connected to one end of the sample, generating a longitudinal temperature difference $\Delta T_x = T_1 - T_2$ (Fig. 1a). The other end of the sample was attached to a heat sink with silver paint. The copper block is used as heat sink, which does not result in any detectable contamination of the thermal Hall signal (more details are available in Supplementary Note 4). The thermal conductivity along the $x$ axis was calculated as $\kappa_{xx} = (Q/\Delta T_x)(L/wt)$. For the samples of $SiO_2$, $x$ is along the [001] axis. For the samples of $SrTiO_3$, $KTaO_3$, $MgO$, $MgAl_2O_4$, Si and Ge, $x$ is along the [100] axis. By applying a magnetic field $H$ along the $-x$, $y$, and $z$ directions, respectively, a transverse gradient $\Delta T_y = T_3 - T_2$ was generated (Fig. 1a). The thermal Hall conductivity was defined as $\kappa_{xy} = \kappa_{yy}(\Delta T_y/\Delta T_x)(L/w)$, where $\kappa_{yy}$ is the longitudinal thermal conductivity along the $y$ axis. We assumed $\kappa_{yy} = \kappa_{xx}$ for crystals with high symmetry ($SrTiO_3$, $KTaO_3$, $MgO$, $MgAl_2O_4$, Si and Ge), while for $SiO_2$ with a chiral crystal structure, $\kappa_{yy}$ has been measured (Supplemental Fig. S6). For the measurements with in-plane fields, the samples are carefully mounted to the heat sink. The large values of the planar $\kappa_{xy}$ in $SiO_2$ exclude the possibility that the planar thermal Hall signals come from the conventional $\kappa_{xy}$ due to the misalignment of the magnetic field. The contamination from $\kappa_{xx}$ in $\kappa_{xy}$ due to contact misalignment is removed by performing field anti-symmetrization of the transverse temperature difference $\Delta T_y$, i.e. $\Delta T_y = [\Delta T_y(H) - \Delta T_y(-H)]/2$. The positive directions of magnetic field are illustrated by color arrows in Fig. 1a.

# References

1. Onose, Y. et al. Observation of the magnon Hall effect. *Science* **329**, 297-299 (2010).
2. Hirschberger, M., Chisnell, R., Lee, Y. S. & Ong, N. P. Thermal Hall effect of spin excitations in a Kagome magnet. *Phys. Rev. Lett.* **115**, 106603 (2015).
3. Zhang, H. et al. Anomalous Thermal Hall effect in an insulating van der Waals magnet. *Phys. Rev. Lett.* **127**, 247202 (2021).
4. Hirschberger, M., Krizan, J. W., Cava, R. J. & Ong, N. P. Large thermal Hall conductivity of neutral spin excitations in a frustrated quantum magnet. *Science* **348**, 106-109 (2015).
5. Kasahara, Y. et al. Majorana quantization and half-integer thermal quantum Hall effect in a Kitaev spin liquid. *Nature* **559**, 227-231 (2018).

6. Strohm, C., Rikken, G. L. J. A. & Wyder, P. Phenomenological evidence for the phonon Hall effect. *Phys. Rev. Lett.* **95**, 155901 (2005).
7. Grissonnanche, G. et al. Giant thermal Hall conductivity in the pseudogap phase of cuprate superconductors. *Nature* **571**, 376-380 (2019).
8. Ideue, T., Kurumaji, T., Ishiwata, S. & Tokura, Y. Giant thermal Hall effect in multiferroics. *Nat. Mater.* **16**, 797-802 (2017).
9. Chen, L., Boulanger, M.-E., Wang, Z.-C., Tafti, F. & Taillefer, L. Large phonon thermal Hall conductivity in the antiferromagnetic insulator $Cu_3TeO_6$. *Proc. Natl. Acad. Sci. U.S.A.* **119**, e2208016119 (2022).
10. Li, X., Fauqué, B., Zhu, Z. & Behnia, K. Phonon thermal Hall effect in strontium titanate. *Phys. Rev. Lett.* **124**, 105901 (2020).
11. Li, X. et al. The phonon thermal Hall angle in black phosphorus. *Nat. Commun.* **14**, 1027 (2023).
12. Yokoi, T. et al. Half-integer quantized anomalous thermal Hall effect in the Kitaev material candidate α-$RuCl_3$. *Science* **373**, 568–572 (2021).
13. Imamura, K. et al. Majorana-fermion origin of the planar thermal Hall effect in the Kitaev magnet α-$RuCl_3$. *Sci. Adv.* **10**, eadk3539 (2024).
14. Grissonnanche, G. et al. Chiral phonons in the pseudogap phase of cuprates. *Nat. Phys.* **16**, 1108-1111 (2020).
15. Boulanger, M.-E. et al. Thermal Hall conductivity in the cuprate Mott insulators $Nd_2CuO_4$ and $Sr_2CuO_2Cl_2$. *Nat. Commun.* **11**, 5325 (2020).
16. Ataei, A. et al. Phonon chirality from impurity scattering in the antiferromagnetic phase of $Sr_2IrO_4$. *Nat. Phys.* (2024).
17. Lefrançois, É. et al. Evidence of a phonon Hall effect in the Kitaev spin liquid candidate α-$RuCl_3$. *Phys. Rev. X* **12**, 021025 (2022).
18. Gillig, M. et al. Phononic-magnetic dichotomy of the thermal Hall effect in the Kitaev material $Na_2Co_2TeO_6$. *Phys. Rev. Res.* **5**, 043110 (2023).
19. Uehara, T., Ohtsuki, T., Udagawa, M., Nakatsuji, S. & Machida, Y. Phonon thermal Hall effect in a metallic spin ice. *Nat. Commun.* **13**, 4604 (2022).
20. Xu, C. et al. Thermal Hall effect in the van der Waals ferromagnet $CrI_3$. *Phys. Rev. B* **109**, 094415 (2024).
21. Vallipuram, A. et al. Role of magnetic ions in the thermal Hall effect of the paramagnetic insulator $TmVO_4$. *arXiv*: 2310.10643 (2023).
22. Chen, L. et al. Planar thermal Hall effect from phonons in cuprates. *arXiv*: 2310.07696 (2023).
23. Chen, L. et al. Planar thermal Hall effect from phonons in a Kitaev candidate material. *Nat. Commun.* **15**, 3513 (2024).
24. Nagaosa, N., Sinova, J., Onoda, S., MacDonald, A. H. & Ong, N. P. Anomalous Hall effect. *Rev. Mod. Phys.* **82**, 1539-1592 (2010).
25. Laughlin, R. B. Anomalous auantum Hall effect: An incompressible quantum fluid with fractionally charged excitations. *Phys. Rev. Lett.* **50**, 1395-1398 (1983).
26. von Klitzing, K. The quantized Hall effect. *Rev. Mod. Phys.* **58**, 519-531 (1986).
27. Novoselov, K. S. et al. Room-temperature quantum Hall effect in graphene. *Science* **315**, 1379-1379 (2007).

28. Kasahara, Y. et al. Unusual thermal Hall effect in a Kitaev spin liquid candidate α-RuCl$_3$. *Phys. Rev. Lett.* **120**, 217205 (2018).
29. Mori, M., Spencer-Smith, A., Sushkov, O. P. & Maekawa, S. Origin of the phonon Hall effect in rare-earth garnets. *Phys. Rev. Lett.* **113**, 265901 (2014).
30. Zhang, Y., Teng, Y., Samajdar, R., Sachdev, S. & Scheurer, M. S. Phonon Hall viscosity from phonon-spinon interactions. *Phys. Rev. B* **104**, 035103 (2021).
31. Samajdar, R., Chatterjee, S., Sachdev, S. & Scheurer, M. S. Thermal Hall effect in square-lattice spin liquids: A Schwinger boson mean-field study. *Phys. Rev. B* **99**, 165126 (2019).
32. Mangeolle, L., Balents, L. & Savary, L. Phonon thermal Hall conductivity from scattering with collective fluctuations. *Phys. Rev. X* **12**, 041031 (2022).
33. Mangeolle, L., Balents, L. & Savary, L. Thermal conductivity and theory of inelastic scattering of phonons by collective fluctuations. *Phys. Rev. B* **106**, 245139 (2022).
34. Guo, H., Joshi, D. G. & Sachdev, S. Resonant thermal Hall effect of phonons coupled to dynamical defects. *Proc. Natl. Acad. Sci. U.S.A.* **119**, e2215141119 (2022).
35. Ueda, H. et al. Chiral phonons in quartz probed by X-rays. *Nature* **618**, 946-950 (2023).
36. Flebus, B. & MacDonald, A. H. Phonon Hall viscosity of ionic crystals. *Phys. Rev. Lett.* **131**, 236301 (2023).
37. Onoda, S., Sugimoto, N. & Nagaosa, N. Intrinsic versus extrinsic anomalous Hall effect in ferromagnets. *Phys. Rev. Lett.* **97**, 126602 (2006).
38. Ceresoli, D., Marchetti, R. & Tosatti, E. Electron-corrected Lorentz forces in solids and molecules in a magnetic field. *Phys. Rev. B* **75**, 161101 (2007).
39. Saito, T., Misaki, K., Ishizuka, H. & Nagaosa, N. Berry phase of phonons and thermal Hall effect in nonmagnetic insulators. *Phys. Rev. Lett.* **123**, 255901 (2019).
40. Chen, J.-Y., Kivelson, S. A. & Sun, X.-Q. Enhanced thermal Hall effect in nearly ferroelectric insulators. *Phys. Rev. Lett.* **124**, 167601 (2020).
41. Qin, T., Zhou, J. & Shi, J. Berry curvature and the phonon Hall effect. *Phys. Rev. B* **86**, 104305 (2012).
42. Czajka, P. et al. Planar thermal Hall effect of topological bosons in the Kitaev magnet α-RuCl$_3$. *Nat. Mater.* **22**, 36-41 (2022).
43. Yao, W. et al. Topological spin excitations in a three-dimensional antiferromagnet. *Nat. Phys.* **14**, 1011-1015 (2018).
44. Tian, S. et al. Ferromagnetic van der Waals Crystal VI$_3$. *J. Am. Chem. Soc.* **141**, 5326-5333 (2019).

## Acknowledgements

We thank Prof. Yuanbo Zhang's group for providing us a piece of intrinsic semiconductor Si single crystal and Prof. Rui Peng for providing us a piece of KTaO$_3$ single crystal. We also thank Prof. Yuan Li's group for providing us the single crystals of $Cu_3TeO_6$, and Prof. Hechang Lei's group for providing us the single crystals of $VI_3$, to validate our experiment set-up. This work is supported by the Natural Science Foundation of China (Grant No. 12174064), the Shanghai Municipal Science and Technology Major Project (Grant No. 2019SHZDZX01), and the National Key Research and Development Program of China (Grant No. 2022YFA1402203).

## Author Contributions

S.Y.L. and X.Z. conceived the idea and designed the experiments. X.B.J. and X.Z. performed the thermal transport measurements with the help from W.B.W., H.R.W., and Y.H.J. X.B.J., X.Z. and S.Y.L. analyzed the data. X.Z., X.B.J. and S.Y.L. wrote the paper with assistance from all the authors. X.B.J., X.Z. contributed equally to this work.

## Competing interests

The authors declare no competing interests.

## Additional Information

**Supplementary information** is available for this paper at URL inserted when published.

**Correspondence** and requests for materials should be addressed to S.Y.L. (shiyan_li@fudan.edu.cn) and X.Z. (xuzhang_fd@fudan.edu.cn).

Figure 1

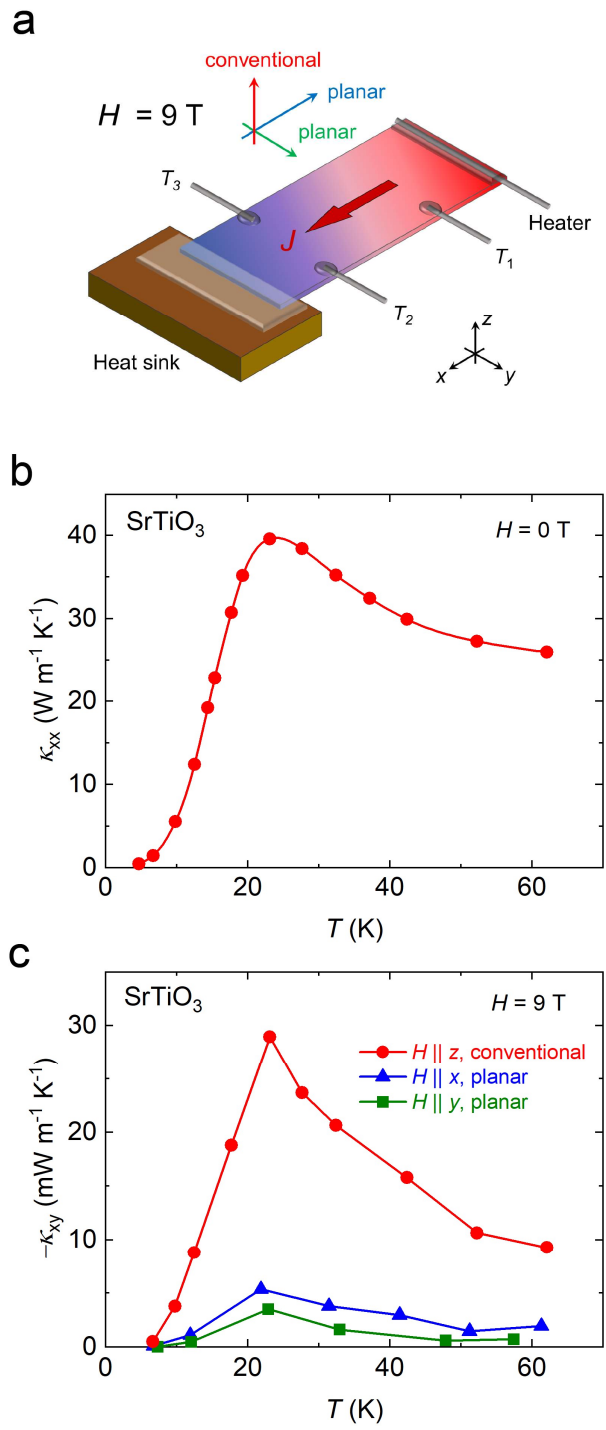

Figure 2

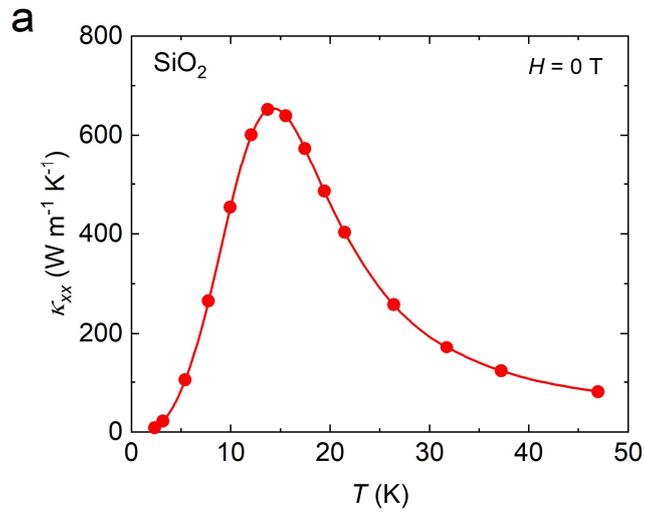

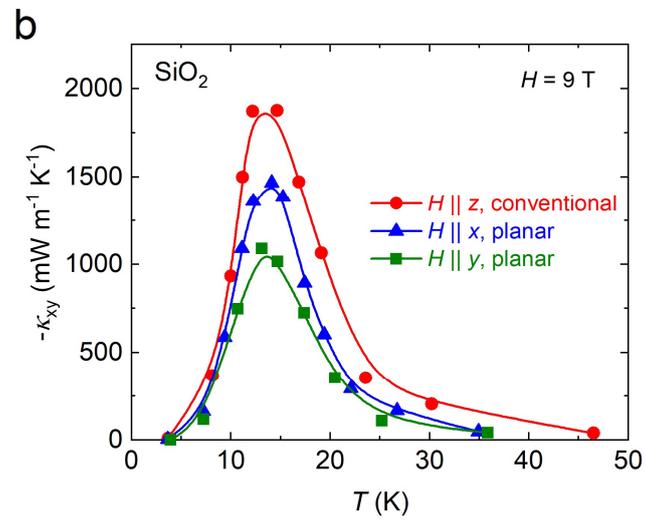

Figure 3

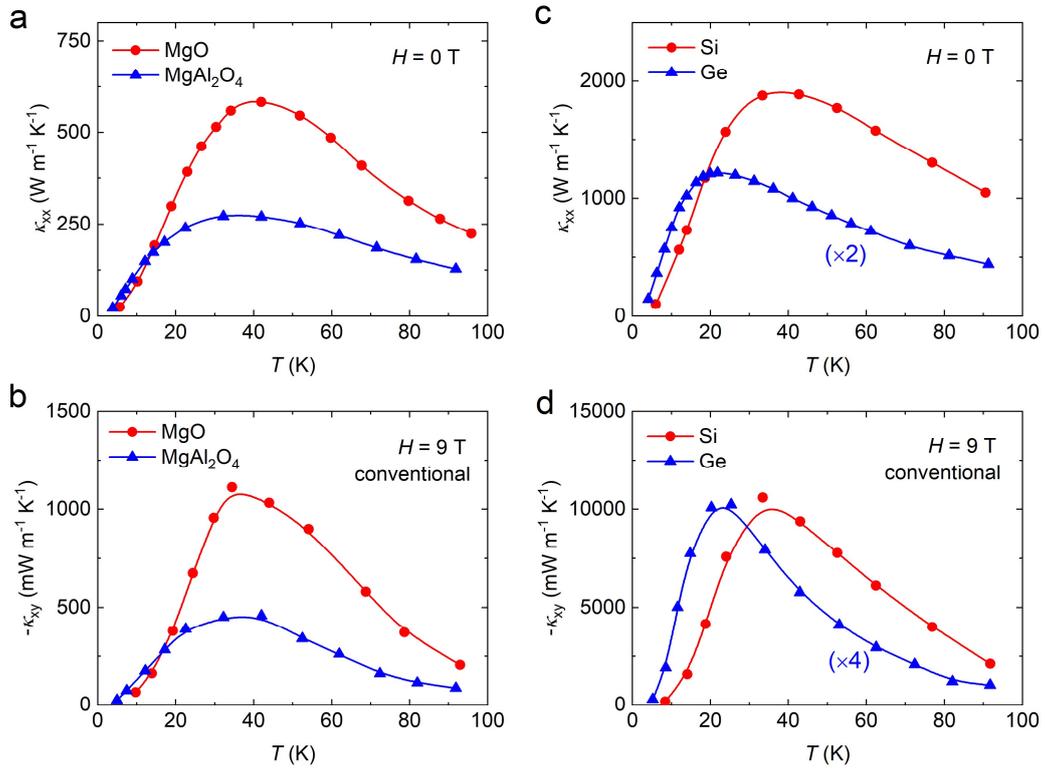

Figure 4

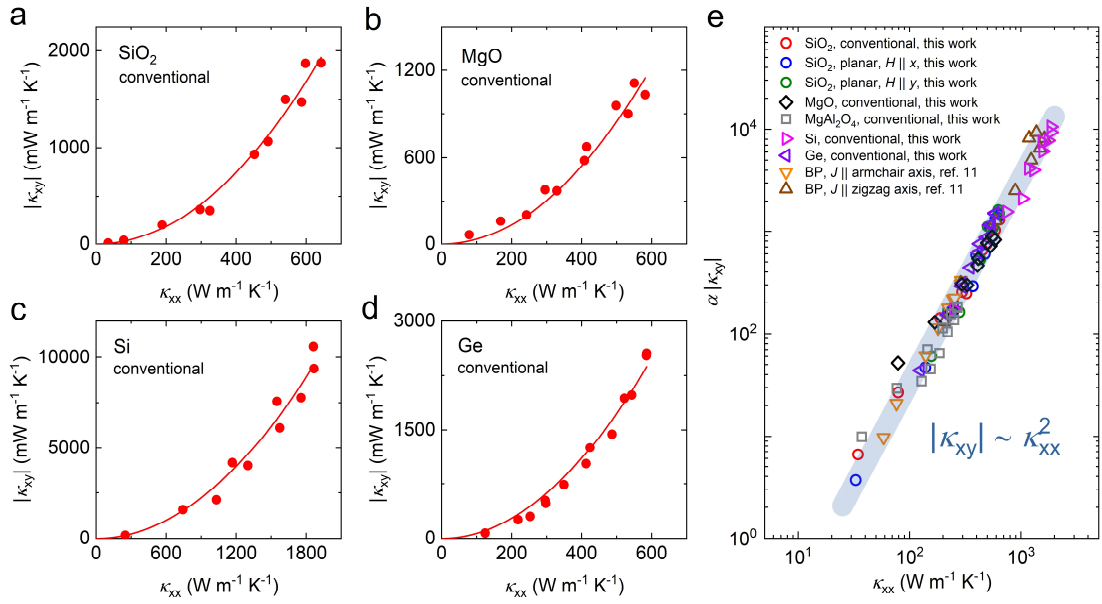

**Figure captions**

**Figure 1 | Experiment setup and phonon thermal transport in SrTiO₃. a**, Schematic of the thermal transport experiment setup. The thermal current $J$ is applied along $x$. The magnetic fields along three directions give the conventional configuration ($H \parallel z$) and two planar configurations ($H \parallel x$ and $H \parallel y$), which are shown with colored arrows. **b**, Thermal conductivity $\kappa_{xx}$ of SrTiO₃ single crystal as a function of temperature in zero field. **c**, Thermal Hall conductivity $\kappa_{xy}$ of SrTiO₃ measured at $H = 9$ T in different configurations as shown in **a**, plotted as $-\kappa_{xy}$ vs $T$. The existence of $\kappa_{xy}$ in two planar configurations is evident.

**Figure 2 | Phonon thermal transport in SiO₂. a**, Thermal conductivity of SiO₂ in zero field, plotted as $\kappa_{xx}$ vs $T$. It shows a phonon peak around 15 K. **b**, Thermal Hall conductivity $\kappa_{xy}$ of SiO₂ measured at $H = 9$ T in conventional and planar configurations shown in Fig. 1a, plotted as $-\kappa_{xy}$ vs $T$. The large $\kappa_{xy}(T)$ in all configurations closely follows $\kappa_{xx}(T)$.

**Figure 3 | Phonon thermal transport in MgO, MgAl₂O₄, Si and Ge. a**, $\kappa_{xx}$ of MgO and MgAl₂O₄ as a function of temperature in zero field. **b**, $\kappa_{xy}$ of MgO and MgAl₂O₄ measured at $H = 9$ T in the conventional configuration, plotted as $-\kappa_{xy}$ vs $T$. **c** and **d,** The same plots for $\kappa_{xx}$ ($H = 0$ T) and $\kappa_{xy}$ ($H = 9$ T) in the case of Si and Ge single crystals. All samples show large $\kappa_{xy}(T)$, following the temperature dependence of $\kappa_{xx}(T)$.

**Figure 4 | Scaling law of phonon THE in non-magnetic insulators and semiconductors. a**, **b**, **c**, **d** $|\kappa_{xy}|$ plotted as a function of $\kappa_{xx}$ for SiO₂, MgO, Si and Ge. The experimental data can be fitted by a function of $|\kappa_{xy}| = A\kappa_{xx}^2$ (the red line). **e**, The scaling law for non-magnetic insulators and semiconductors. To facilitate a better comparison, the $\kappa_{xx}$ of these insulators and semiconductors is multiplied by a factor $\alpha$.

Supplementary Information for

# "Discovery of universal phonon thermal Hall effect in crystals"

Xiaobo Jin[1*], Xu Zhang[1*✉], Wenbo Wan[1], Hanru Wang[1], Yihan Jiao[1] and Shiyan Li[1,2,3✉]

[1]*State Key Laboratory of Surface Physics, Department of Physics, Fudan University, Shanghai 200433, China*

[2]*Shanghai Research Center for Quantum Sciences, Shanghai 201315, China*

[3]*Collaborative Innovation Center of Advanced Microstructures, Nanjing 210093, China*

Corresponding author. Email: shiyan_li@fudan.edu.cn (S.Y.L.); xuzhang_fd@fudan.edu.cn (X.Z.)

# Supplementary Note 1: Thermal Hall measurements on glass and $Cu_3TeO_6$

To validate the thermal Hall results from our experiment setup, we first carried out measurements on an amorphous non-magnetic sample—a piece of glass, in which there is negligible thermal Hall signal[1]. A standard microscope slide was cut and polished to the similar dimensions as other samples. Figure S1a shows the thermal conductivity of the glass sample. Indeed, the thermal Hall signal in glass is negligible, as shown in Fig. S1b, which is consistent with previous report[1]. This excludes an artificial finite $\kappa_{xy}$ from our experiment setup.

For comparison, we measured the antiferromagnetic insulator $Cu_3TeO_6$, in which large $\kappa_{xy}$ has been reported[2]. As expected, we observed a large thermal Hall signal in $Cu_3TeO_6$ with a thermal Hall angle of 0.25% at 9 T (Fig. S1b), which is comparable with that in $SiO_2$. Further, we did detailed measurements on the $Cu_3TeO_6$ sample. In Fig. S1c, $\kappa_{xx}(T)$ shows a phonon peak at around 20 K and a kink at $T_N$, suggesting the high quality of our sample. In a field of 12 T, large $\kappa_{xy}(T)$ is observed, following the trend of $\kappa_{xx}(T)$. All these behaviors are consistent with previous report, depicted as blue circles[2]. Note that the relatively smaller value of $\kappa_{xy}$ in our sample is due to the smaller $\kappa_{xx}$ and magnetic field. The reproducibility of these data validates our experiment setup.

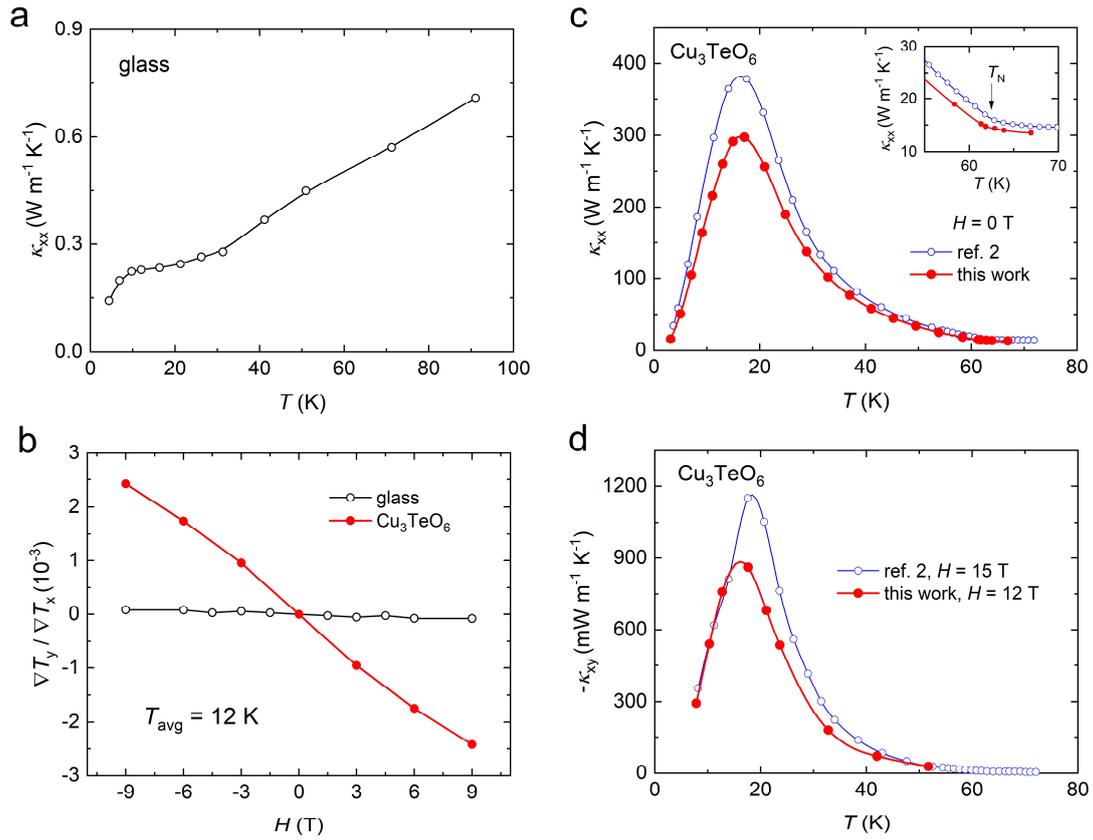

**Figure S1 | The thermal Hall signals in glass and Cu₃TeO₆. a**, $\kappa_{xx}$ of glass as a function of temperature in zero field. **b**, The thermal Hall angles of glass and Cu₃TeO₆ as a function of magnetic field. The average temperature $T_{avg} = (T_1 + T_2)/2$ was hold at 12 K for both samples. **c**, $\kappa_{xx}$ of Cu₃TeO₆ as a function of temperature in zero field. The inset gives the zoom-in $\kappa_{xx}(T)$ around $T_N$. **d**, $\kappa_{xy}$ of Cu₃TeO₆ measured at $H = 12$ T as a function of temperature. For comparison, the data from ref. 2 were also plotted in **c** and **d** as blue empty circles.

## Supplementary Note 2: Thermal Hall measurements on VI$_3$ and CrI$_3$

If $\kappa_{xy}(T)$ always follows $\kappa_{xx}(T)$ in our samples, we still worry about whether it is an artificial result. Therefore, we chose to measure ferromagnetic insulators VI$_3$ and CrI$_3$, since the $\kappa_{xy}(T)$ of VI$_3$ has special temperature dependence and different sign from that of CrI$_3$ (refs. 3 and 4). Figure S2a and b show $\kappa_{xx}(T)$ and $\kappa_{xy}(T)$ of VI$_3$, respectively. $\kappa_{xx}(T)$ manifests a kink at $T_c$ = 50 K. A large positive $\kappa_{xy}$, as reported in ref. 3, arises around $T_c$, then shows a broad peak at low temperature. For CrI$_3$, as shown in Fig. S2c and d, it has a negative $\kappa_{xy}$ and it peaks at the same temperature as $\kappa_{xx}(T)$. Compared with the data extracted from refs. 3 and 4, all the features we observed in VI$_3$ and CrI$_3$ are consistent with previous reports[3,4], thus further validate the reliability of our thermal Hall measurements.

Starting from a universal phonon THE in crystals, the conclusion of this work, one can see that the magnetic environment below $T_c$ = 61 K in CrI$_3$ only slightly affects the phonon $\kappa_{xy}(T)$. However, for VI$_3$, the distinct shape and positive sign of $\kappa_{xy}(T)$ suggest that the magnetic environment below $T_c$ = 50 K has huge effect on the phonon THE. We discuss it in the main text together with other ferromagnetic insulators.

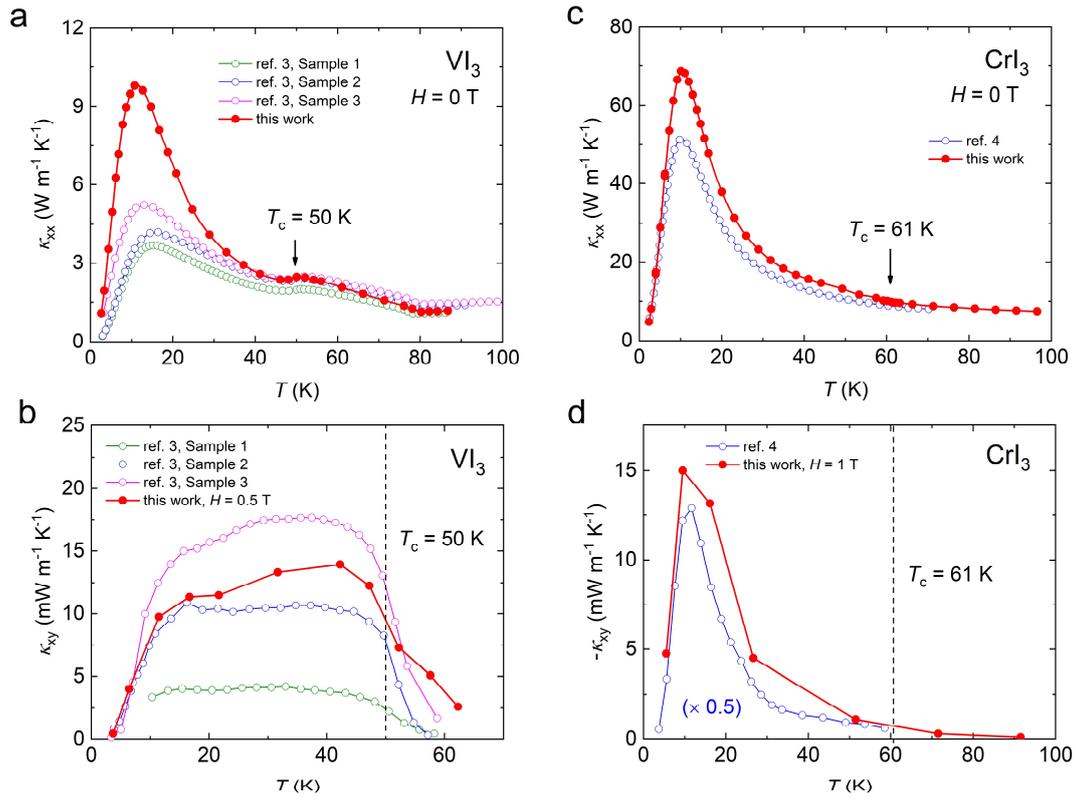

**Figure S2 | The thermal Hall effect in VI$_3$ and CrI$_3$. a**, $\kappa_{xx}$ of VI$_3$ as a function of temperature in zero field. **b**, $\kappa_{xy}$ of VI$_3$ measured at $H$ = 0.5 T in the spin polarized state, plotted as $\kappa_{xy}$ vs $T$. For comparison, the data from ref. 3 were also plotted in **a** and **b** as empty circles. **c**, $\kappa_{xx}$ of CrI$_3$ as a function of temperature in zero field. **d**, $\kappa_{xy}$ of CrI$_3$ measured at $H$ = 1 T in the spin polarized state, plotted as $\kappa_{xy}$ vs $T$. For comparison, the data from ref. 4 were also plotted in **c** and **d** as blue empty circles.

## Supplementary Note 3: Crystal structure of SiO$_2$

SiO$_2$ (quartz) has a chiral crystal structure, in which SiO$_4$ tetrahedra form a chiral helix along [001]. The resulting chiral space group is either P3$_2$21 (left quartz) or P3$_1$21 (right quartz). Figure S3 gives the crystal structure of right quartz.

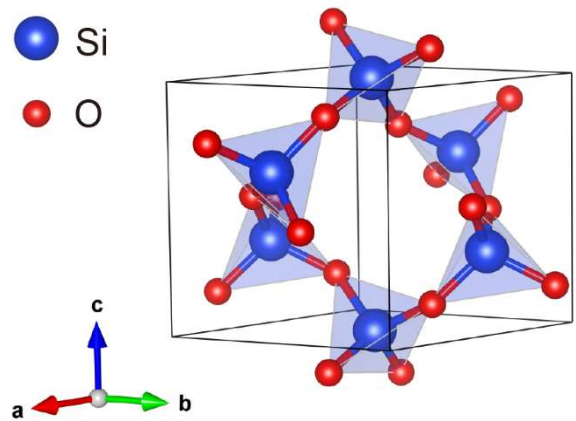

**Figure S3 | Crystal structure of SiO$_2$.**

## Supplementary Note 4: Check the effect of copper block heat sink

In our thermal transport experiments, a copper block is used as the heat sink. In ref. 5, the authors observed a non-zero Hall signal in non-magnetic $Y_2Ti_2O_7$ (the data was extracted and plotted in Fig. S4a) and attributed this signal to temperature gradients on the heat sink (brass block). Although it is unlikely to generate a finite transverse temperature gradient that could affect the sample due to the large size of the copper block, we conducted comparative experiment to check it. The measurement on $SiO_2$ sample was conducted again with a different configuration. The measurement in the main text involved the direct adhesion of the sample to the copper block (Fig. 1a), while the comparative experiment took another configuration where the sample was connected to the copper block via silver wire (Configuration B in Fig. S4b). The results of both measurements are consistent (Fig. S4c and d), confirming that using copper block as the heat sink does not result in any detectable contamination on thermal Hall data. Based on our finding, the non-zero thermal Hall signal observed in ref. 5 should arise from $Y_2Ti_2O_7$ itself, in which the thermal Hall angle is approximately $0.25 \times 10^{-3}$ at 15 K and 9 T, on the same order of magnitude as that of $SrTiO_3$. In ref. 5, the thermal Hall angle of magnetic $Tb_2Ti_2O_7$ is approximately $8 \times 10^{-3}$ at 15 K and 9 T, which is comparable to the value of $5.5 \times 10^{-3}$ for Si at 30 K and 9 T. The finite thermal Hall signal of non-magnetic $Y_2Ti_2O_7$ observed in ref. 5 further confirms the universal phonon THE in crystals discovered in our current work.

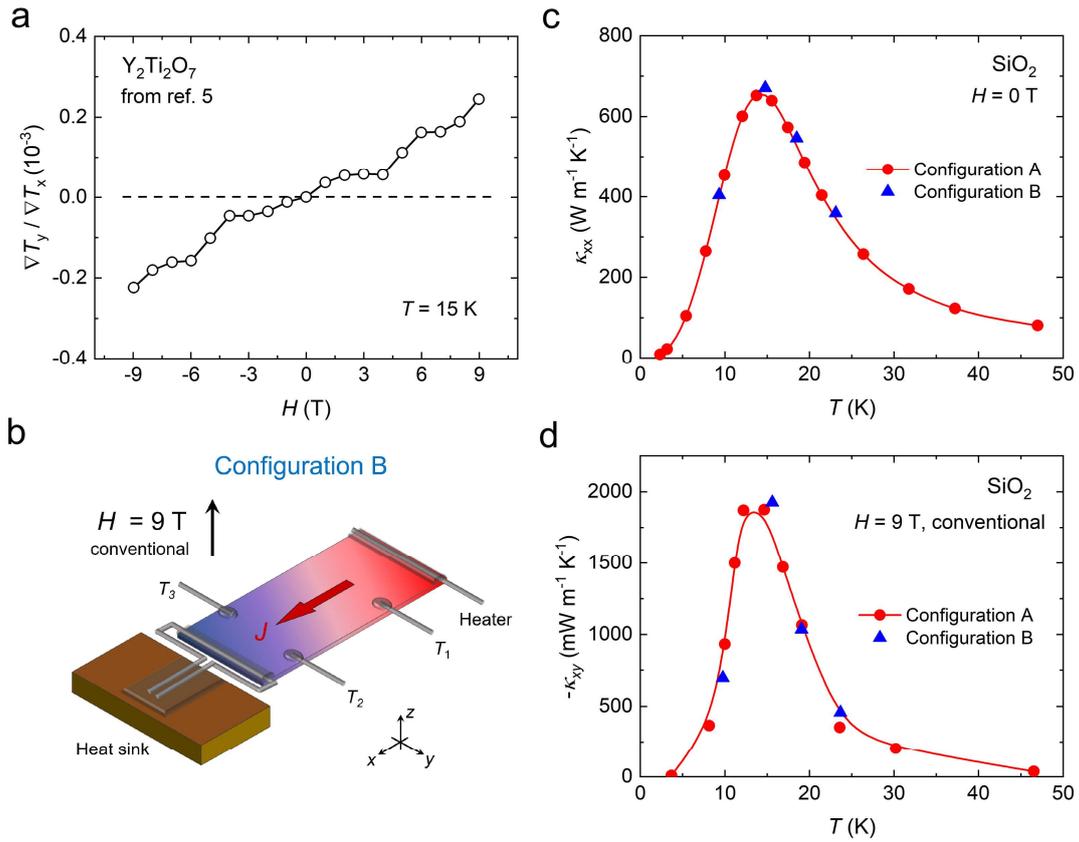

**Figure S4 | Check the effect of copper block heat sink. a**, The thermal Hall angle of $Y_2Ti_2O_7$ as a function of magnetic field. The data are from ref. 5. **b**, A measurement configuration in which the sample is connected to the copper block via silver wire. This is named Configuration B to distinguish it from the configuration in main text (Configuration A). **c**, **d**, Comparison of two thermal transport measurements with Configuration A and B on the same $SiO_2$ sample. The results of both measurements are consistent, confirming that using copper block as the heat sink does not result in any detectable contamination on thermal Hall data.

## Supplementary Note 5: Thermal Hall measurements on KTaO$_3$

Previous work suggests that KTaO$_3$ (KTO) has a tiny $\kappa_{xy}$ (ref. 6), which is at odds with the universal phonon THE observed in this study. Therefore, we also conducted thermal Hall measurements on KTO single crystal. As shown in Fig. S5a, the temperature dependence of $\kappa_{xx}$ for our sample is consistent with that reported in ref. 7, and slightly different from that in ref. 6 at low temperature. Clearly, a finite $\kappa_{xy}$ is observed, with the peak value about one-third of STO (Fig. S5). This finding further confirms the existence of universal phonon THE in crystals. The difference between our data and the previous work on KTO needs further study.

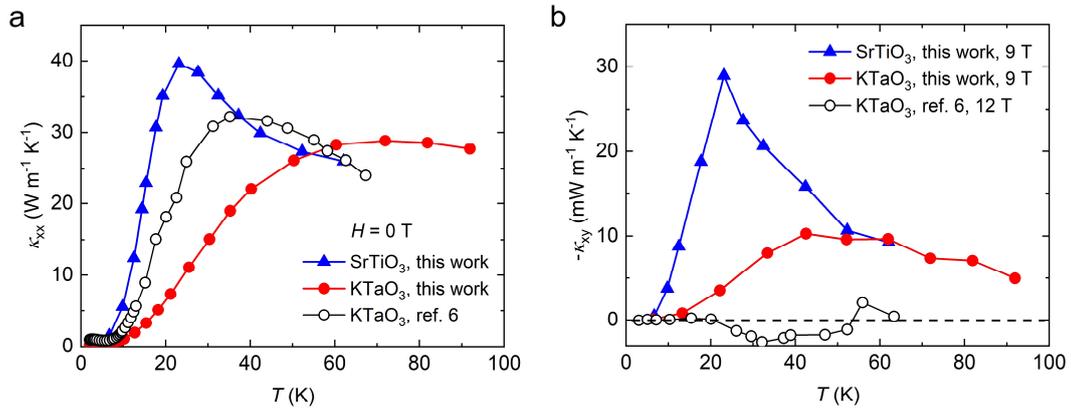

**Figure S5 | The thermal Hall effect in KTaO$_3$ and SrTiO$_3$. a**, $\kappa_{xx}$ of KTaO$_3$ and SrTiO$_3$ measured in this work as a function of temperature in zero field. **b**, $\kappa_{xy}$ of KTaO$_3$ and SrTiO$_3$ measured at $H$ = 9 T as a function of temperature. The $\kappa_{xx}$ and $\kappa_{xy}$ of KTaO$_3$ from ref. 6 are also plotted for comparison.

## Supplementary Note 6: $\kappa_{yy}$ of SiO$_2$ single crystals

The thermal Hall conductivity is defined as $\kappa_{xy} = \kappa_{yy}(\Delta T_y/\Delta T_x)(L/w)$, where $\kappa_{yy}$ is the longitudinal thermal conductivity along the *y* axis. For crystals with high symmetry (SrTiO$_3$, KTaO$_3$, MgO, MgAl$_2$O$_4$, Si and Ge), *x* is along the [100] axis, and [100], [010] and [001] directions are equivalent. Therefore, we take $\kappa_{yy} = \kappa_{xx}$ when calculating $\kappa_{xy}$. For the samples of SiO$_2$, *x* is along the [001] axis and *y* is along the [1$\bar{1}$0] axis. Due to the chiral crystal structure, $\kappa_{yy}$ and $\kappa_{xx}$ in SiO$_2$ should be unequal. We cut and polished another SiO$_2$ sample with the longest edges along [1$\bar{1}$0] axis (*y*) to the same dimensions as the sample in main text. $\kappa_{yy}$ was measured in this sample with $J \parallel [1\bar{1}0]$. As shown in Fig. S6, $\kappa_{yy}$ exhibits a slightly larger value than $\kappa_{xx}$.

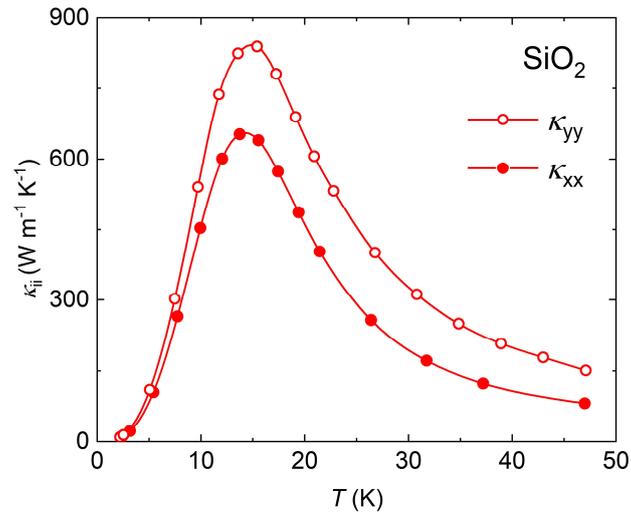

**Figure S6 | $\kappa_{xx}$ and $\kappa_{yy}$ of SiO$_2$ single crystals.**

# Supplementary Note 7: Scaling law in MgAl$_2$O$_4$ and black phosphorous

In Fig. S7, we plot the conventional $\kappa_{xy}$ ($\kappa_{xz}$ and $\kappa_{zx}$) as a function of $\kappa_{xx}$ for MgAl$_2$O$_4$ and black phosphorous (BP). The data of BP were extracted from ref. 8. The thermal transport data of MgAl$_2$O$_4$ and BP also obey the scaling law $|\kappa_{xy}| \sim \kappa_{xx}^2$, as the other non-magnetic insulators and semiconductors studied in this work.

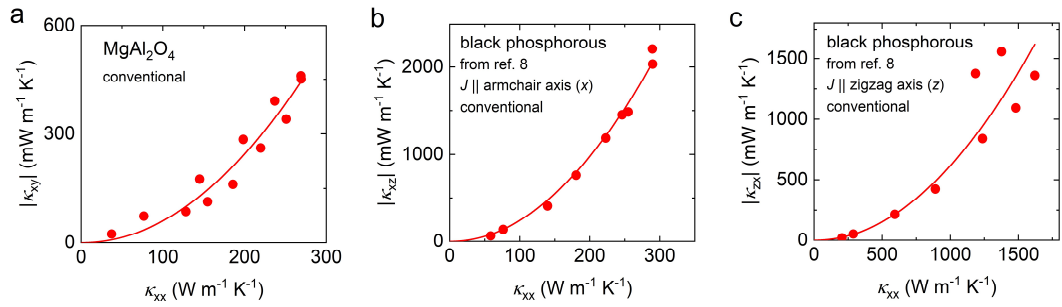

**Figure S7 | $|\kappa_{xy}|$ vs $\kappa_{xx}$ for MgAl$_2$O$_4$ and black phosphorous.**

## Supplementary Note 8: Scaling law between planar $\kappa_{xy}$ and $\kappa_{xx}$ in SiO$_2$

In Fig. S8, we plot the conventional $\kappa_{xy}$ and two planar $\kappa_{xy}$ as a function of $\kappa_{xx}$ for SiO$_2$ sample. One can clearly see that the planar $\kappa_{xy}$ also obeys the scaling law $|\kappa_{xy}| \sim \kappa_{xx}^2$, as the conventional $\kappa_{xy}$ does.

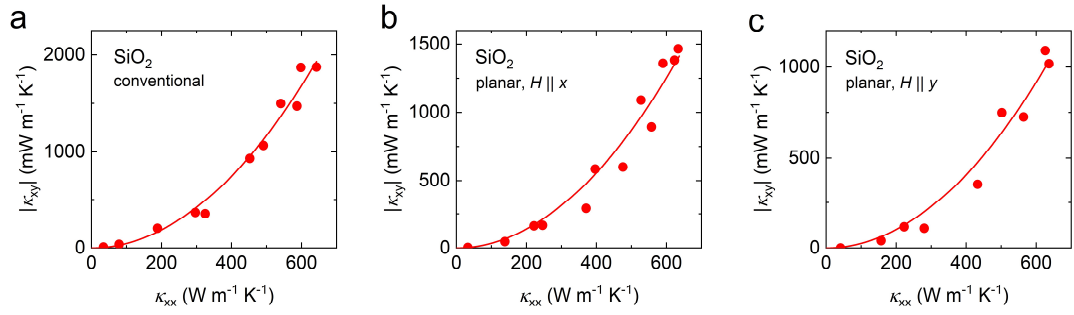

**Figure S8 | $|\kappa_{xy}|$ vs $\kappa_{xx}$ in different configurations for SiO$_2$.**

# Supplementary Note 9: Absence of scaling law between $\kappa_{xy}$ and $\kappa_{xx}$ in magnetic materials

In Fig. S9, we plot $|\kappa_{xy}|$ vs. $\kappa_{xx}$ for various magnetic insulators in which THE has been observed[9-12]. Unlike most non-magnetic insulators and semiconductors measured in this work, these materials manifest a deviation from the scaling law $|\kappa_{xy}| \sim \kappa_{xx}^2$, indicating the effect of magnetic environment on the phonon THE.

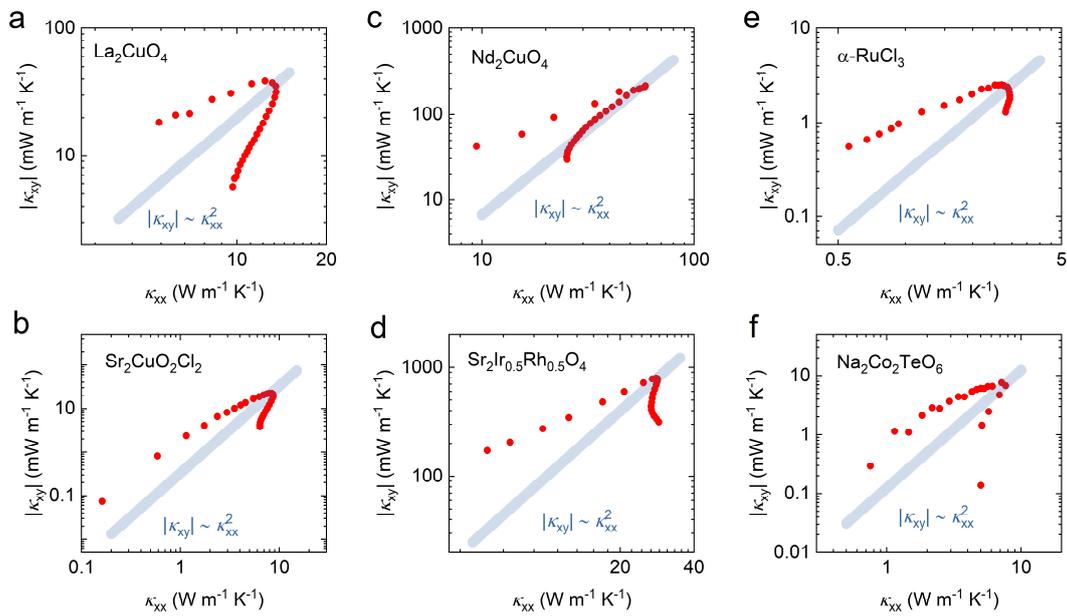

**Figure S9 | $|\kappa_{xy}|$ vs $\kappa_{xx}$ for various magnetic insulators.** $|\kappa_{xy}|$ plotted as a function of $\kappa_{xx}$ for $La_2CuO_4$ (ref. 9), $Sr_2CuO_2Cl_2$ (ref. 9), $Nd_2CuO_4$ (ref. 9), $Sr_2Ir_{0.5}Rh_{0.5}O_4$ (ref. 10), $\alpha$-$RuCl_3$ (ref. 11) and $Na_2Co_2TeO_6$ (ref. 12). The blue lines represent the scaling relation of $|\kappa_{xy}| \sim \kappa_{xx}^2$.

# Supplementary references

1. Kim, H.-L. et al. Modular thermal Hall effect measurement setup for fast-turnaround screening of materials over wide temperature range using capacitive thermometry. *Rev. Sci. Instrum.* **90**, 103904 (2019).
2. Chen, L., Boulanger, M.-E., Wang, Z.-C., Tafti, F. & Taillefer, L. Large phonon thermal Hall conductivity in the antiferromagnetic insulator $Cu_3TeO_6$. *Proc. Natl. Acad. Sci. U.S.A.* **119**, e2208016119 (2022).
3. Zhang, H. et al. Anomalous thermal Hall effect in an insulating van der Waals magnet. *Phys. Rev. Lett.* **127**, 247202 (2021).
4. Xu, C. et al. Thermal Hall effect in the van der Waals ferromagnet $CrI_3$. *Phys. Rev. B* **109**, 094415 (2024).
5. Hirschberger, M., et al. Large thermal Hall conductivity of neutral spin excitations in a frustrated quantum magnet. *Science* **348**, 106-109 (2015)
6. Li, X., Fauqué, B., Zhu, Z. & Behnia, K. Phonon thermal Hall effect in strontium titanate. *Phys. Rev. Lett.* **124**, 105901 (2020).
7. Tachibana, M., Kolodiazhnyi, T., Takayama-Muromachi, E. Thermal conductivity of perovskite ferroelectrics. *Appl. Phys. Lett*. **93**, 092902 (2008).
8. Li, X. et al. The phonon thermal Hall angle in black phosphorus. *Nat. Commun.* **14**, 1027 (2023).
9. Boulanger, M.-E. et al. Thermal Hall conductivity in the cuprate Mott insulators $Nd_2CuO_4$ and $Sr_2CuO_2Cl_2$. *Nat. Commun.* **11**, 5325 (2020).
10. Ataei, A. et al. Phonon chirality from impurity scattering in the antiferromagnetic phase of $Sr_2IrO_4$. *Nat. Phys.* (2024).
11. Lefrançois, É. et al. Evidence of a phonon Hall effect in the Kitaev spin liquid candidate α-$RuCl_3$. *Phys. Rev. X* **12**, 021025 (2022).
12. Yang, H. et al. Significant thermal Hall effect in the 3d cobalt Kitaev system $Na_2Co_2TeO_6$. *Phys. Rev. B* **106**, L081116 (2022).